\documentclass[epj]{webofc}
\usepackage[varg]{txfonts}   % Web of Conferences font
\usepackage{longtable}
\usepackage{amsmath}
\usepackage{amssymb}
\usepackage{graphicx}
\usepackage{bm}% bold math
\usepackage{footmisc}
\usepackage{afterpage}
\usepackage{float}
\usepackage{lscape}
\usepackage{longtable}
\usepackage{float}

\wocname{EPJ Web of Conferences}

\woctitle{CONF12}

\newcommand*{\no}{\noindent}
\newcommand*{\bea}{\begin{eqnarray}}
\newcommand*{\eea}{\end{eqnarray}}
\newcommand*{\be}{\begin{equation}}
\newcommand*{\ee}{\end{equation}}
\newcommand*{\pd}{\partial}
\newcommand*{\pdm}{\pd_{\mu}}

\newcommand*{\pref}[1]{(\ref{#1})}

\newcommand*{\mn}{{\mu\nu}}

\newcommand*{\nn}{\nonumber}

\newcommand{\bma}{\begin{pmatrix}}
\newcommand{\ema}{\end{pmatrix}}

\begin{document}
\selectlanguage{english}
\title{Gauge engineering and propagators}

\author{Axel Maas\inst{1}\fnsep\thanks{\email{axel.maas@uni-graz.at}}
}

\institute{Institute of Physics, NAWI Graz, University of Graz,
Universit\"atsplatz 5, A-8010 Graz, Austria}

\abstract{Beyond perturbation theory gauge-fixing becomes more involved due to the Gribov-Singer ambiguity: The appearance of additional gauge copies requires to define a procedure how to handle them. For the case of Landau gauge the structure and properties of these additional gauge copies will be investigated. Based on these properties gauge conditions are constructed to account for these gauge copies.

The dependence of the propagators on the choice of these complete gauge-fixings will then be investigated using lattice gauge theory for Yang-Mills theory. It is found that the implications for the infrared, and to some extent mid-momentum behavior, can be substantial. In going beyond the Yang-Mills case it turns out that the influence of matter can generally not be neglected. This will be briefly discussed for various types of matter.}

\maketitle

\section{Introduction}

The determination of the gauge-dependent correlation functions, especially propagators and vertices, is a long-standing challenge \cite{Alkofer:2000wg,Fischer:2006ub,Binosi:2009qm,Maas:2011se,Boucaud:2011ug,Vandersickel:2012tg}. One of the main problems is that the concept of a gauge itself becomes more involved beyond perturbation theory. The reason for this is the appearance of the Gribov-Singer ambiguity, essentially the presence of perturbatively absent additional gauge copies. This obstructs any perturbative gauge-fixing from being complete, i.\ e.\ able to fully specify how the gauge orbit should be sampled \cite{Gribov:1977wm,Singer:1978dk}. Thus, extended sampling prescriptions are necessary to provide well-defined gauges.

While formulating complete gauge fixings is possible in principle, the remaining practical obstacle is to find such prescriptions which can be implemented in different methods. The motivation for such comparisons of gauge-dependent, rather than gauge-invariant, quantities is that none of the available non-perturbative methods is truly exact. A comparison of results at the level of gauge-dependent quantities has turned out to be quite useful to identify artifacts due to approximations \cite{Maas:2011se}. Reporting some preliminary results for steps along the way to this goal is the aim here. This extends previous results provided in \cite{Maas:2009se,Maas:2011ba,Maas:2011se,Maas:2013vd} and will be completed elsewhere \cite{Maas:unpublished}.

The problem of constructing a complete gauge-fixing usually boils down to the fact that after perturbative gauge-fixing a set of Gribov copies remains. It needs to be specified how to treat them, i.\ e.\ how to sample the residual gauge orbit. This is done by introducing some weight function with which to average over this residual gauge orbit. This has been investigated using anything from an average over the full residual gauge orbit \cite{Hirschfeld:1978yq,vonSmekal:2008en,Fischer:2008uz,Maas:2012ct}, a subset of the residual gauge orbit \cite{Maas:2011se,Maas:2013vd,vonSmekal:2008en,vonSmekal:2008es,Parrinello:1990pm,Serreau:2013ila} to a $\delta$-function-like weight \cite{Cucchieri:1997dx,Vandersickel:2012tg,Boucaud:2011ug,Bornyakov:2010nc,Fachin:1991pu,Fischer:2008uz,Maas:2009se,Maas:2008ri,Silva:2004bv,Schaden:2014bea,Sternbeck:2012mf,Zwanziger:1993dh,Henty:1996kv}.

Here, the case of a class of gauges averaging over a part of the residual gauge orbit will be analyzed, namely averaging over the so-called first Gribov region, to be defined below. The prime interest is, whether this has any (sizable) influence on the correlation functions, here the gluon and ghost propagators, as well as the running coupling derived from them in the miniMOM scheme \cite{vonSmekal:2009ae}. The reason for this interest is that these correlation functions can be calculated with various methods \cite{Alkofer:2000wg,Fischer:2006ub,Binosi:2009qm,Maas:2011se,Boucaud:2011ug,Vandersickel:2012tg}. If they are sensitive to the choice of gauge, they can be used as a marker to compare if the implementation of the same gauge fixings in different methods agree.

The design of the gauges utilizes knowledge of the structure of the first Gribov region obtained from \cite{Maas:2015nva}. All of this will be done using lattice gauge theory, in which the implementation of such gauges is straight-forward, as discussed in \cite{Maas:2011se}. The technical details of these simulations will be presented elsewhere \cite{Maas:unpublished}, but essentially follow \cite{Maas:2009se,Maas:2015nva}.

Of course, if the outlined program is eventually successful, this can be turned into a feature: By choosing a suitable gauge, the correlation functions can be engineered such that calculations become simpler. This idea, which is also implemented in perturbation theory, is behind the term gauge engineering.

\section{Defining the gauges}

The gauges to be investigated here are defined in a three-step process. The first step is to implement the perturbative Landau gauge \cite{Maas:2011se}. This creates a hypersurface in the space of gauge-orbits, which cuts every gauge orbit such that no second cut occurs for any infinitesimal gauge transformation. Thus, the residual gauge orbits are a set of discrete gauge copies on the gauge orbit. These gauge copies, the so-called Gribov copies, are then classified according to the number of negative eigenvalues of the Faddeev-Popov operator. Of these, only those copies are retained which have no negative eigenvalues, which make up the so-called first Gribov region \cite{Vandersickel:2012tg}. See \cite{Maas:2015nva} for details of how this is done in the lattice calculations presented here. Finally, the remaining Gribov copies will be weighted with the weight function
\be
w(\xi,\zeta)=\exp\left({\cal N}+\frac{\xi}{V}\int d^dxd^dy\pdm^x\bar{c}^a(x)\pdm^yc^a(y)-\frac{\zeta}{V}\int d^dx A_\mu^a A_\mu^a\right),\label{wf}
\ee
\no where ${\cal N}$ is a normalization factor, $A_\mu$ are the gluon fields, $c$ and $\bar{c}$ are the ghost fields, and $\xi$ and $\zeta$ are additional gauge parameters The gauges studied in \cite{Maas:2011se,Maas:2013vd,vonSmekal:2008en,vonSmekal:2008es,Parrinello:1990pm,Serreau:2013ila,Cucchieri:1997dx,Vandersickel:2012tg,Boucaud:2011ug,Bornyakov:2010nc,Fachin:1991pu,Fischer:2008uz,Maas:2009se,Maas:2008ri,Silva:2004bv,Schaden:2014bea,Sternbeck:2012mf,Zwanziger:1993dh,Henty:1996kv} all correspond to particular values of both gauge parameters. The most well-known gauges of these are the minimal Landau gauge at $\xi=\zeta=0$, the absolute Landau gauge at $\xi=0$ and $\zeta=-\infty$ and the extreme Landau $b$ gauges at $\xi=\pm\infty$ and $\zeta=0$ \cite{Maas:2011se,Maas:2011ba,Maas:2009se,Maas:unpublished}.

In practical lattice calculations of these gauges there are two caveats. One is that it is not possible to include all Gribov copies, as it is numerically not possible to obtain all, even if a constructive way of generating them would be available, which is not. In fact, already to differentiate between different Gribov copies is a non-trivial issue \cite{Maas:2015nva}. Thus, especially for large absolute values of the gauge parameters, tails are usually not adequately sampled. Thus, any results on the gauge dependence can be at most a lower limit, as will be discussed in more detail in section \ref{s:impact}. The sample of Gribov copies used here is obtained by the method described in \cite{Maas:2015nva}. The second caveat is that any given gauge-fixing algorithm could introduce an additional, algorithmic bias. Though results so far do not show any indications that this is the case \cite{Maas:2011ba,Maas:2013vd,Maas:unpublished}, there is no proof.

As it turns out that finite values of $\zeta$ and $\xi$ smoothly interpolate between the results obtained from sending the gauge parameters to $\pm\infty$ \cite{Maas:unpublished,Maas:2013vd}, here only the extreme cases, as well as $\zeta=\xi=0$, will be studied.

\begin{figure}
\includegraphics[width=\linewidth]{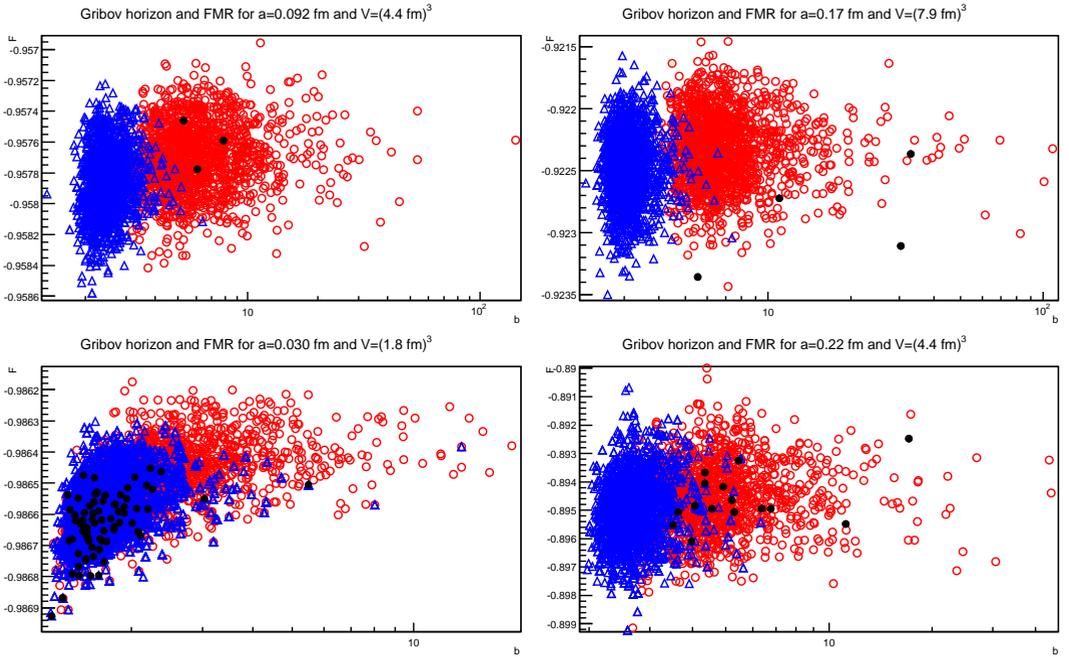}
\caption{\label{fig:gribov} The first Gribov region in three dimensions for different lattice volumes and discretizations. Open blue triangles have the smallest values of $F$ found, the so-called fundamental modular region (FMR), and red open circles the largest value of $b$, each found on their respective gauges orbits. Black full circles are Gribov copies satisfying both criteria. See \cite{Maas:2015nva} for more details. Results are in three dimensions, using the noted lattice parameters.}
\end{figure}

It is interesting to compare the layout of the Gribov region in comparison to how it is sampled. For this purpose, the Gribov copies sampled at some of the extremes of the gauge parameters are plotted in terms of their weight functions, abbreviated as
\bea
b&=&\int d^dxd^dy\pdm^x\bar{c}^a(x)\pdm^yc^a(y)\nn\\
F&=&-1+\frac{\zeta}{V}\int d^dx A_\mu^a A_\mu^a\nn,
\eea
\no in figure \ref{fig:gribov}. Shown are those Gribov copies which minimize $F$ and maximize $b$. Both areas tend to more strongly decompose the larger the physical volume. Thus, indeed, gauges triggering to these extreme values will sample different parts of the first Gribov region. On the other hand, they tend to wash out the dependence on the other coordinate, as they are, more or less, equally distributed.

\section{Impact on the correlation functions}\label{s:impact}

The gauge-dependent quantities, which arguably have been studied most, are the gluon and ghost propagators \cite{Alkofer:2000wg,Fischer:2006ub,Binosi:2009qm,Maas:2011se,Boucaud:2011ug,Vandersickel:2012tg}. It is also these quantities which will be considered here. How they are determined for any given Gribov copy can be found in \cite{Cucchieri:2006tf}. As only the low-momentum behavior will be studied, in all cases the momenta are chosen along an axis. To obtain the final result, they will be averaged over the Gribov copies obtained in the lattice simulations, weighted by \pref{wf}. After that, they will be averaged over configurations as described in \cite{Cucchieri:2006tf}. As here only the extreme gauges are considered, this is equivalent to choosing only the Gribov copy which has the most extreme corresponding value or, for the minimal Landau gauge, a random copy. In fact, finite values of the gauge parameter only interpolate between these choices \cite{Maas:unpublished,Maas:2013vd}.

The running coupling in the miniMOM scheme \cite{vonSmekal:2009ae} can be obtained from these correlation functions directly, as this coupling is given by
\be
\alpha(p^2)=\frac{\alpha(\mu^2)}{(d-1)(N_c^2-1)^3}p^6P_\mn D_\mn^{aa}(p^2,\mu^2)(D_G^{aa}(p^2,\mu^2))^2\nn,
\ee
\no where $\mu^2$ is the renormalization point, $d$ the dimensionality, $N_c$ is the number of colors, $P_\mn$ is the usual transverse projector, $D_\mn^{ab}$ is the gluon propagator, and $D_G^{ab}$ is the ghost propagator. The dependence on the renormalization drops out in this process. As it is a product, it is particularly sensitive to the modifications of the correlation functions, and therefore is a suitable indication for the severity of the effect. Moreover, the running coupling is an important ingredient in many approximation schemes for hadronic physics \cite{Alkofer:2000wg,Fischer:2006ub}, though dependencies on its low-momentum behavior for hadronic observables appear to be rather small \cite{Blank:2010pa,Costa:2010pp}. This is as it should be, given the result below that this behavior is, in fact, strongly affected by the gauge choice.

As it turns out \cite{Maas:2015nva}, the averaging is hampered by the fact that the number of Gribov copies found is even on moderately (physically) sized lattices much smaller than the actual number of Gribov copies, by orders of magnitude. Thus, any numerical result can be considered to be at most a lower limit to the actual sensitivity to Gribov copies and thus the gauge choice. To illustrate this, in the following also the dependency of the correlation function at the lowest momentum on the number of sampled Gribov copies is determined.

\begin{figure}
 \includegraphics[width=\textwidth]{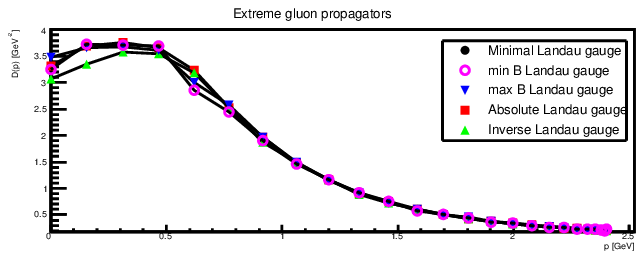}
 \includegraphics[width=\textwidth]{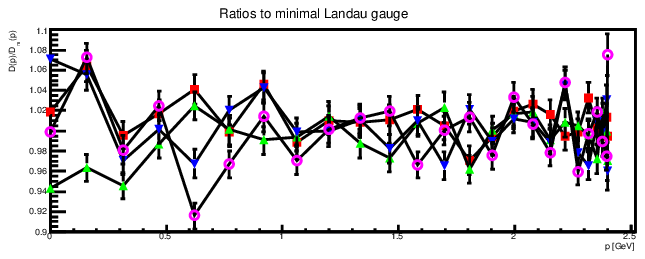}
 \includegraphics[width=\textwidth]{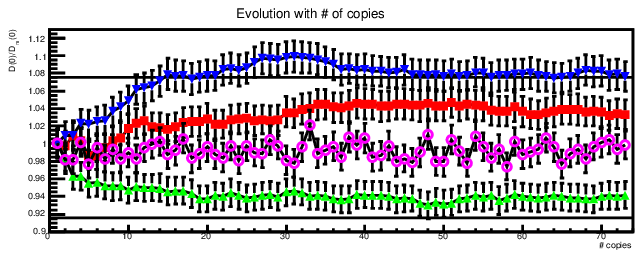}
 \caption{\label{fig:gp}The gluon propagator for various gauges (top panel) and the ratio to its value in minimal Landau gauge (middle panel). The bottom panel shows the evolution of the ratio at zero momentum as a function of included (genuine \cite{Maas:2015nva}) Gribov copies. Results are in three dimension with a lattice spacing $a^{-1}=1.20$ GeV and a volume of $(Na)^3=(48a)^3=(7.9$ fm$)^3$. Full lines in the bottom panel are from extrapolations in the number of Gribov copies, see text.}
\end{figure}

\begin{figure}
 \includegraphics[width=\textwidth]{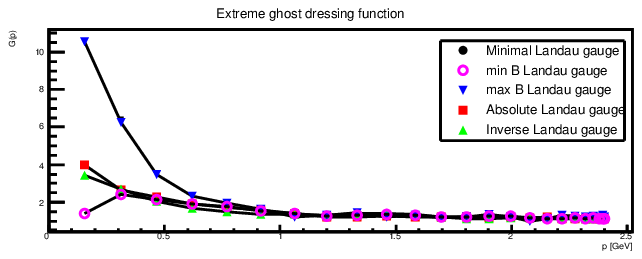}
 \includegraphics[width=\textwidth]{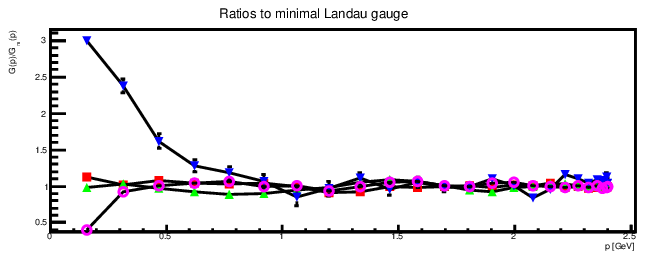}
 \includegraphics[width=\textwidth]{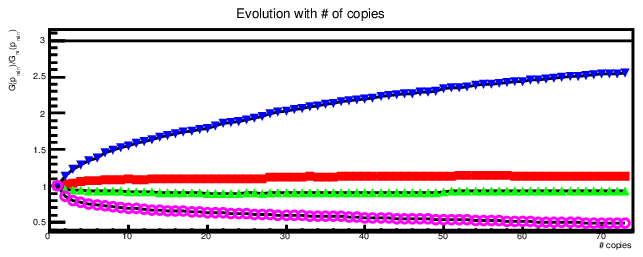}
 \caption{\label{fig:ghp}The ghost dressing function for various gauges (top panel) and the ratio to its value in minimal Landau gauge (middle panel). The bottom panel shows the evolution of the ratio at the lowest possible momentum on this lattice, 157 MeV, as a function of included (genuine \cite{Maas:2015nva}) Gribov copies. Results are in three dimension with a lattice spacing $a^{-1}=1.20$ GeV and a volume of $(Na)^3=(48a)^3=(7.9$ fm$)^3$. Full lines in the bottom panel are from extrapolations in the number of Gribov copies, see text.}
\end{figure}

\begin{figure}
 \includegraphics[width=\textwidth]{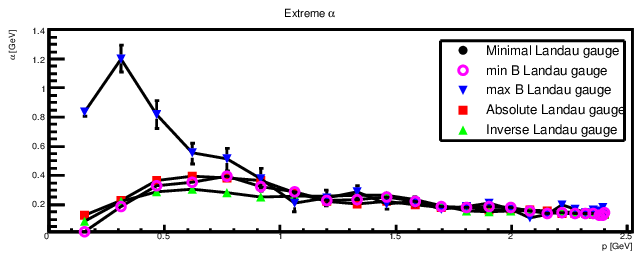}
 \includegraphics[width=\textwidth]{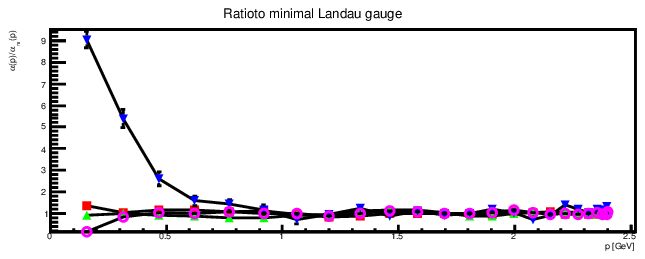}
 \includegraphics[width=\textwidth]{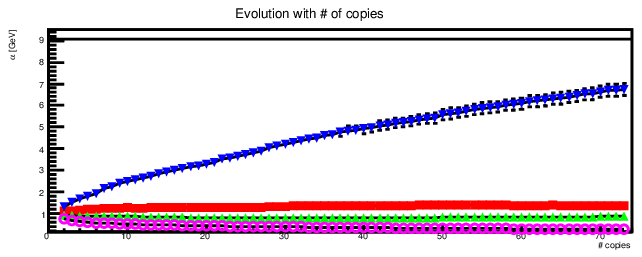}
 \caption{\label{fig:alpha}The ghost dressing function for various gauges (top panel) and the ratio to its value in minimal Landau gauge (middle panel). The bottom panel shows the evolution of the ratio at the lowest possible momentum on this lattice, 157 MeV, as a function of included (genuine \cite{Maas:2015nva}) Gribov copies. Results are in three dimension with a lattice spacing $a^{-1}=1.20$ GeV and a volume of $(Na)^3=(48a)^3=(7.9$ fm$)^3$. Full lines in the bottom panel are from extrapolations in the number of Gribov copies, see text.}
\end{figure}

The results are shown in figures \ref{fig:gp}-\ref{fig:alpha} for the gluon, the ghost, and the running coupling, respectively. The gluon propagator shows essentially no statistically significant influence on any gauge choice. Its values also converges as a function of Gribov copies very quickly to its (presumed) limit of all Gribov copies included. The limit is actually estimated from an ansatz $c+a/N_g$ of the corresponding quantity in the number of Gribov copies, averaging over several choices of fitting range \cite{Maas:unpublished}. This weak dependence is in accordance with other investigations of some of the presented gauge choices \cite{Maas:2011se,Maas:2013vd,Cucchieri:1997dx,Boucaud:2011ug,Bornyakov:2010nc,Maas:2009se,Maas:2008ri,Silva:2004bv,Sternbeck:2012mf}.

The situation is very different for the ghost. The impact strongly depends on the choice of gauge. Choices, which utilize the quantity $F$ show, however, also very little change, again in accordance with previous investigations \cite{Maas:2011se,Maas:2013vd,Cucchieri:1997dx,Boucaud:2011ug,Bornyakov:2010nc,Maas:2009se,Maas:2008ri,Silva:2004bv,Sternbeck:2012mf}. In this case also the dependence on the number of Gribov copies is mild. This drastically changes when studying gauges based on the quantity $b$. Here, the impact is already at about half a GeV statistically significant. This is even more pronounced in the case of the running coupling. This is not surprising, as the gluon propagator essentially does not change, and there is hence nothing to offset the squared effect from the ghost. This has also corresponding implications for the dependence on the number of included Gribov copies.

However, most notable is that there appears to be nonetheless no qualitative change, as was originally hoped for \cite{Maas:2009se,Maas:2008ri} - eventually deep enough in the infrared the qualitative behavior changes always to the one in minimal Landau gauge, in accordance with the expectations of \cite{Fischer:2008uz}. Only reaching this point requires more and more effort. Thus, it appears that all possibilities to select Gribov copies in the first Gribov region studied so far lead to the same qualitative, but not quantitative, behavior. The latter is at least true for all lattice volumes studied so far for this range of gauges \cite{Maas:2011se,Maas:2009se,Maas:unpublished, Sternbeck:2012mf}.

\section{Summary and the role of matter}

The bottom line of this investigation is that some correlation functions, most notably the ghost propagator and quantities derived from it, depend strongly on the choice of gauge for momenta at or below roughly 500 MeV. However, this behavior is only quantitative in all cases studied. Furthermore, an estimate of the precise size of this quantitative effect is obstructed by the quick rise in the number of Gribov copies with volume. Nonetheless, this implies that a comparison of results from different methods makes for some quantities only sense if the same gauge is chosen. Especially for the comparison of lattice and functional methods, this implies that still some effort needs to be invested to have full control over the implementation of the same gauge \cite{Maas:2009se,Maas:2011se,Fischer:2008uz}. Still, expressions like \pref{wf} are already a step towards a continuum formulation, as they no longer make explicit reference to Gribov copies, but only to fields, and thus are easier to handle in continuum formulations. It is also encouraging that these formulations give the same result as the ones based on the individual manipulation of Gribov copies \cite{Maas:2013vd,Maas:unpublished}.

A last issue concerns the influence of matter on the results presented here. The gauge conditions employed here, the (non-aligned \cite{Maas:2012ct}) Landau gauges are well-defined in the presence of any matter fields. However, they never include the matter fields explicitly. Especially, the presence of matter fields cannot turn a gauge copy in Landau gauge into something else or remove it from the gauge orbit. Thus, matter fields can at most give different residual gauge orbits different weights in the path integral.

While this subject has not yet been studied in great detail, it appears that matter in QCD-like situations does not have any significant impact on this question \cite{Maas:2010nc,Maas:2011jf}. However, this drastically changes when a Brout-Englert-Higgs effect is at work \cite{Maas:2010nc}: In marked contrast to the situation here, the standard algorithms do find no Gribov copies \cite{Maas:2010nc}. Whether this is an algorithmic deficiency or whether indeed the residual gauge orbits in this case have a different number of Gribov copies is unclear. At any rate, the corresponding implications are far reaching, and deserve a better understanding. This is especially true, as also continuum investigations support a change of behavior \cite{Capri:2013oja}. In this respect, studies of the superconformal case, as a third possibility, may also be useful, as also in this case a different behavior is motivated by continuum investigations \cite{Capri:2014tta,Maas:2015zjx}.

\bibliography{bib}

\begin{thebibliography}{38}

\bibitem{Alkofer:2000wg}
R.~Alkofer, L.~von Smekal, Phys. Rept. \textbf{353}, 281 (2001),  (2001),
  \texttt{hep-ph/0007355}

\bibitem{Fischer:2006ub}
C.S. Fischer, J. Phys. \textbf{G32}, R253 (2006),  (2006),
  \texttt{hep-ph/0605173}

\bibitem{Binosi:2009qm}
D.~Binosi, J.~Papavassiliou, Phys. Rept. \textbf{479}, 1 (2009),  (2009),
  \texttt{0909.2536}

\bibitem{Maas:2011se}
A.~Maas, Phys. Rep. \textbf{524}, 203 (2013),  (2013), \texttt{1106.3942}

\bibitem{Boucaud:2011ug}
P.~Boucaud et~al., Few Body Syst. \textbf{53}, 387 (2012),  (2012),
  \texttt{1109.1936}

\bibitem{Vandersickel:2012tg}
N.~Vandersickel, D.~Zwanziger, Phys.Rept. \textbf{520}, 175 (2012),  (2012),
  \texttt{1202.1491}

\bibitem{Gribov:1977wm}
V.N. Gribov, Nucl. Phys. \textbf{B139}, 1 (1978),  (1978)

\bibitem{Singer:1978dk}
I.M. Singer, Commun. Math. Phys. \textbf{60}, 7 (1978),  (1978)

\bibitem{Maas:2009se}
A.~Maas, Phys. Lett. \textbf{B689}, 107 (2010),  (2010), \texttt{0907.5185}

\bibitem{Maas:2011ba}
A.~Maas, PoS \textbf{QCD-TNT-II}, 028 (2011),  (2011), \texttt{1111.5457}

\bibitem{Maas:2013vd}
A.~Maas, PoS \textbf{ConfinementX}, 034 (2012),  (2012), \texttt{1301.2965}

\bibitem{Maas:unpublished}
A.~Maas (unpublished),  (unpublished)

\bibitem{Hirschfeld:1978yq}
P.~Hirschfeld, Nucl. Phys. \textbf{B157}, 37 (1979),  (1979)

\bibitem{vonSmekal:2008en}
L.~von Smekal, M.~Ghiotti, A.G. Williams, Phys. Rev. \textbf{D78}, 085016
  (2008),  (2008), \texttt{0807.0480}

\bibitem{Fischer:2008uz}
C.S. Fischer, A.~Maas, J.M. Pawlowski, Annals Phys. \textbf{324}, 2408 (2009),
  (2009), \texttt{0810.1987}

\bibitem{Maas:2012ct}
A.~Maas, Mod. Phys. Lett. \textbf{A27}, 1250222 (2012),  (2012)

\bibitem{vonSmekal:2008es}
L.~von Smekal, A.~Jorkowski, D.~Mehta, A.~Sternbeck, PoS \textbf{CONFINEMENT8},
  048 (2008),  (2008), \texttt{0812.2992}

\bibitem{Parrinello:1990pm}
C.~Parrinello, G.~Jona-Lasinio, Phys.Lett. \textbf{B251}, 175 (1990),  (1990)

\bibitem{Serreau:2013ila}
J.~Serreau, M.~Tissier, A.~Tresmontant, Phys.Rev. \textbf{D89}, 125019 (2014),
  (2014), \texttt{1307.6019}

\bibitem{Cucchieri:1997dx}
A.~Cucchieri, Nucl. Phys. \textbf{B508}, 353 (1997),  (1997),
  \texttt{hep-lat/9705005}

\bibitem{Bornyakov:2010nc}
V.G. Bornyakov, V.K. Mitrjushkin, Phys.Rev. \textbf{D84}, 094503 (2011),
  (2011), \texttt{1011.4790}

\bibitem{Fachin:1991pu}
S.~Fachin, C.~Parrinello, Phys.Rev. \textbf{D44}, 2558 (1991),  (1991)

\bibitem{Maas:2008ri}
A.~Maas, Phys. Rev. \textbf{D79}, 014505 (2009),  (2009), \texttt{0808.3047}

\bibitem{Silva:2004bv}
P.J. Silva, O.~Oliveira, Nucl. Phys. \textbf{B690}, 177 (2004),  (2004),
  \texttt{hep-lat/0403026}

\bibitem{Schaden:2014bea}
M.~Schaden, D.~Zwanziger, Phys. Rev. \textbf{D92}, 025001 (2014),  (2014),
  \texttt{1412.4823}

\bibitem{Sternbeck:2012mf}
A.~Sternbeck, M.~M\"uller-Preussker, Phys. Lett. \textbf{B726}, 396 (2012),
  (2012), \texttt{1211.3057}

\bibitem{Zwanziger:1993dh}
D.~Zwanziger, Nucl. Phys. \textbf{B412}, 657 (1994),  (1994)

\bibitem{Henty:1996kv}
D.~Henty, O.~Oliveira, C.~Parrinello, S.~Ryan (UKQCD Collaboration), Phys.Rev.
  \textbf{D54}, 6923 (1996),  (1996), \texttt{hep-lat/9607014}

\bibitem{vonSmekal:2009ae}
L.~von Smekal, K.~Maltman, A.~Sternbeck, Phys. Lett. \textbf{B681}, 336 (2009),
   (2009), \texttt{0903.1696}

\bibitem{Maas:2015nva}
A.~Maas (2015),  (2015), \texttt{1510.08407}

\bibitem{Cucchieri:2006tf}
A.~Cucchieri, A.~Maas, T.~Mendes, Phys. Rev. \textbf{D74}, 014503 (2006),
  (2006), \texttt{hep-lat/0605011}

\bibitem{Blank:2010pa}
M.~Blank, A.~Krassnigg, A.~Maas, Phys. Rev. \textbf{D83}, 034020 (2011),
  (2011), \texttt{1007.3901}

\bibitem{Costa:2010pp}
P.~Costa, O.~Oliveira, P.J. Silva, Phys.Lett. \textbf{B695}, 454 (2011),
  (2011), \texttt{1011.5603}

\bibitem{Maas:2010nc}
A.~Maas, Eur. Phys. J. \textbf{C71}, 1548 (2011),  (2011), \texttt{1007.0729}

\bibitem{Maas:2011jf}
A.~Maas, JHEP \textbf{1105}, 077 (2011),  (2011), \texttt{1102.5023}

\bibitem{Capri:2013oja}
M.~Capri, D.~Dudal, M.~Guimaraes, I.~Justo, S.~Sorella et~al., Annals Phys.
  \textbf{343}, 72 (2013),  (2013), \texttt{1309.1402}

\bibitem{Capri:2014tta}
M.A.L. Capri, M.S. Guimaraes, I.F. Justo, L.F. Palhares, S.P. Sorella, Phys.
  Lett. \textbf{B735}, 277 (2014),  (2014), \texttt{1404.7163}

\bibitem{Maas:2015zjx}
A.~Maas, S.~Zitz, Eur. Phys. J. \textbf{C76}, 113 (2016),  (2016),
  \texttt{1512.06664}

\end{thebibliography}

\end{document}